\documentclass{article}
\usepackage{Format/ijcai20}

\usepackage{times}

\usepackage{soul}
\usepackage{url}
\usepackage[hidelinks]{hyperref}
\usepackage[utf8]{inputenc}
\usepackage[small]{caption}
\usepackage{graphicx}
\usepackage{acro}
\usepackage{amsmath}
\usepackage{amsfonts}
\usepackage{amsthm}
\usepackage{amssymb}
\usepackage{algorithm}
\usepackage{algpseudocode} 
\usepackage[caption=false, font=footnotesize]{subfig}
\usepackage{booktabs}
\usepackage{todonotes}
\usepackage{tikz}
\usetikzlibrary{automata, positioning, arrows, shapes.geometric}

\title{Synthesis of Deceptive Strategies in Reachability Games with Action Misperception}

\author{
Abhishek N. Kulkarni\And
Jie Fu\\
\affiliations
Worcester Polytechnic Institute, Worcester, USA
\emails
\{ankulkarni, jfu2\}@wpi.edu
}

\newif\ifuseboldmathops
\newif\ifuseittextabbrevs
\useboldmathopstrue   % comment out to use mathbb
\useittextabbrevstrue % comment out to use non-italic text abbrevs like e.g.
\ifuseittextabbrevs

	\newcommand{\ie}{{\it i.e.~}}

\else

	\newcommand{\ie}{i.e.~}

\fi

\ifuseboldmathops
\else

\fi

\ifuseboldmathops

\else

\fi

\ifuseboldmathops

\else

\fi

\ifuseboldmathops

\else

\fi

\newcommand{\Eventually}{\Diamond \, }

\newcommand{\dist}[1]{\mathsf{Dist}(#1)}

\newcommand{\supp}{\mathsf{Supp}}

\newtheorem{theorem}{Theorem}
\newtheorem{definition}{Definition}
\newtheorem{example}{Example}
\newtheorem{problem}{Problem}
\newtheorem{lemma}{Lemma}
\newtheorem{assumption}{Assumption}

\newcommand{\win}{\mathsf{Win}}
\newcommand{\occ}{\mathsf{Occ}}
\newcommand{\game}{\mathcal{G}}
\newcommand{\hgame}{\mathcal{H}}

\theoremstyle{plain}

\newtheorem{proposition}{Proposition}
\newtheorem{corollary}{Corollary}[theorem]

\ifdefined \lemma \else
  \newtheorem{lemma}{Lemma}
\fi

\ifdefined \theorem \else
  \newtheorem{theorem}{Theorem}
\fi

\ifdefined \question \else
    \newtheorem{question}{Question}
\fi

\theoremstyle{definition}
\newtheorem{notation}{Notation}

\ifdefined \problem \else
  \newtheorem{problem}{Problem}
\fi

\ifdefined \definition \else
  \newtheorem{definition}{Definition}
\fi

\ifdefined \assumption \else
  \newtheorem{assumption}{Assumption}
\fi

\ifdefined \example \else
  \newtheorem{example}{Example}
\fi

\theoremstyle{remark}

\newcommand{\fig}[1]{\mbox{Fig.~\ref{#1}}}

\newcommand{\sect}[1]{Section~\ref{#1}}

\newcommand{\thm}[1]{Thm.~\ref{#1}}
\newcommand{\lma}[1]{Lemma~\ref{#1}}
\newcommand{\ass}[1]{Assumption~\ref{#1}}

\newcommand{\defn}[1]{Def.~\ref{#1}}
\newcommand{\alg}[1]{Alg.~\ref{#1}}
\newcommand{\prop}[1]{Proposition~\ref{#1}}

\newcommand{\ex}[1]{Example~\ref{#1}}

\DeclareAcronym{ltl}{
	short = LTL, long = Linear Temporal Logic ,
	class = abbrev
}

\DeclareAcronym{scltl}{
	short = sc-LTL, long = Syntactically Co-safe LTL ,
	class = abbrev
}

\DeclareAcronym{dfa}{
	short = DFA, long = Deterministic Finite Automaton ,
	class = abbrev
}

\DeclareAcronym{mdp}{
	short = MDP, long = Markov Decision Process ,
	class = abbrev
}
\newcommand{\act}{{Act}}
\newcommand{\update}{{\eta}}

\newcommand{\dapre}{{\mathsf{DAPre}}}

\DeclareAcronym{sw}{
	short = SW, long = Sure Winning ,
	class = abbrev
}

\DeclareAcronym{asw}{
	short = ASW, long = Almost-Sure Winning ,
	class = abbrev
}

\DeclareAcronym{lsw}{
	short = LSW, long = Limit-Sure Winning ,
	class = abbrev
}

\DeclareAcronym{dasw}{
	short = DASW, long = Deceptive Almost-Sure Winning ,
	class = abbrev
}

\DeclareAcronym{pps}{
	short = PPS, long = Perceptually Permissive Strategy ,
	class = abbrev
}

\DeclareAcronym{rpps}{
	short = RPPS, long = Randomized Perceptually Permissive Strategy ,
	class = abbrev
}

\begin{document}
\maketitle

\begin{abstract}
We consider a class of two-player turn-based zero-sum games on graphs with reachability objectives, known as reachability games, where the objective of Player 1 (P1) is to reach a set of goal states, and that of Player 2 (P2) is to prevent this. In particular, we consider the case where the players have asymmetric information about each other's action capabilities: P2 starts with an incomplete information (misperception) about P1's action set, and updates the misperception when P1 uses an action previously unknown to P2. When P1 is made aware of P2's misperception, the key question is \textit{whether P1 can control P2's perception so as to deceive P2 into selecting actions to P1's advantage?} We show that there might exist a   deceptive winning strategy for P1 that ensures P1's objective is achieved with probability one from a  state otherwise losing for P1, had the information being  symmetric and complete. We present three key results: First, we introduce a dynamic hypergame model to capture the reachability game with evolving misperception of P2. Second, we present a fixed-point algorithm to compute the \ac{dasw} region and \ac{dasw} strategy. Finally, we show that \ac{dasw} strategy is at least as powerful as \ac{asw} strategy in the game in which P1 does not account for P2's misperception. We illustrate our algorithm using a robot motion planning in an adversarial environment.  
\end{abstract}

% =====================================================
% SECTION
% =====================================================
\section{Introduction} \label{sec:introduction}

Synthesis of winning strategies in reachability games is a central problem in   reactive synthesis \cite{Pnueli1989}, control of discrete event systems \cite{Ramadge1989}, and robotics \cite{Fainekos2009}. In a two-player reachability game, a controllable player, P1 (pronoun ``she"), plays against an uncontrollable adversarial player, P2 (pronoun ``he"), to reach the goal states. These games have been extensively studied in algorithmic game theory \cite{deAlfaro2007} and reactive synthesis \cite{Bloem2012}. Polynomial-time algorithms are known for synthesizing \textit{sure-winning} and \textit{almost-sure winning} strategies, when both players have complete and symmetric information. However, the solution concepts for such games under asymmetric information have not been thoroughly studied. 

% The research community is presently interested in exploring the reachability games under partial observability \cite{Ji2018}, asymmetric information \cite{Kulkarni2019}. 

% Recently, the reachability games are being studied in the context of partial observability \cite{Ji2018} and deception \cite{Goes2019}. Also, 

% In recent literature, these games are being studied in the context of partial observability \cite{Ji2018} and synthesis with Linear Temporal Logic under asymmetric information \cite{Kulkarni2019}, bounded synthesis \cite{camacho2018ltl}. 

% \todo[inline]{mention your CDC paper?}
%  \todo[inline]{here you want to relate the reacahbility game to more complex problems: reactive synthesis is LTL cosafe specifications---so that you methods are not too limited.}

% We study two-player turn-based reachability games on graphs under information asymmetry. 
Information asymmetry arises when a player has some private information that are not shared with others \cite{eric1989games}. We consider the case when P1 has complete information about both players' action capabilities, but P2 starts with an incomplete information about P1's action capabilities. As two players interact, their information evolves. Particularly, when P1 uses an action previously unknown to P2, P2 can update his knowledge about the other's capabilities using an \textit{inference-mechanism}. In response, P2 would update his counter-strategy. 
We are interested in the following question:
\textit{if P1 is made aware of the initial information known to P2 and his inference mechanism, can P1 find a strategy to control P2's information in such a way that P2's counter-strategy given his evolving information  is advantageous to P1?} In the context of reachability game, the question translates to: \textit{
 from a state that is losing for P1 in a  game  with symmetric information, 
can P1 reach his goal  from the same state when the information is asymmetric?}
We note that a strategy of P1 that controls P2's information to P1's advantage is indeed  deceptive   \cite{Ettinger2010}. In this paper, we show that such a deceptive winning strategy may exist and propose an algorithm to synthesize it.

% We note that a strategy of P1 that controls P2's information to P1's advantage is indeed a deceptive strategy \cite{Ettinger2010}. Deception is said to be the key indicator of intelligence \cite{Alloway2015} and has been studied in computer science in the domains of artificial intelligence \cite{Turing1950}, game theory \cite{hespanha2000deception}, human-robot interaction \cite{Shim2013,Hancock2011} and cyber-security \cite{Jajodia2016book,Pawlick2019}. 

We approach the question based on the modeling and solution concepts of hypergame \cite{Bennett1977}. A hypergame  allows players to play different games and further allows players to model the games that others are playing. 
In the literature, hypergames and Bayesian games are common models to capture game-theoretic interactions with asymmetric, incomplete information. In Bayesian games, each player uses his incomplete information to define a probability distribution over the possible types of the opponent. The distributions over types are assumed to be common knowledge. In hypergames, no such probabilistic characterizations of incomplete information is used or assumed. For action deception, P2 has incomplete information about P1's capabilities but does not have a prior knowledge about the set of possible types of P1.
Thus, we adopt the hypergame model \cite{gharesifard2012evolution} to understand action deception. In the past, hypergame model has been used to study deception \cite{Gutierrez,kovach2016temporal}. These papers mainly focus on extending the notion of Nash equilibrium to level-$k$ normal form hypergames. \cite{gharesifard2014stealthy} use the notion of H-digraph to establish necessary and sufficient conditions for deceivability. An H-digraph models a hypergame as a graph with nodes representing different outcomes in a normal-form game. However, our game model is not a normal-form game, but instead a game on graph. A hypergame model based on a game on graph has been defined in \cite{Kulkarni2019} where one player has incomplete information about the other's task specification. However, their model assumes that the misperception of P2  remains constant, whereas in our case, the game is dynamic with both players' evolving perception. % is not the case with our problem. 

To synthesize deceptive strategies, we first define a dynamic hypergame on graph and then introduce an algorithm to identify the deceptive almost-sure winning region, which contains a set of states from where P1 has a deceptive strategy to ensure that the goal is reached with probability one. Our main contributions are as follows.
% \todo{DASW region is a complicated term, can you write it out?}

% Therefore, in this work, we start by defining the notion of dynamic hypergame on graph that extends the definition of hypergame on graph in \cite{Kulkarni2019}. The dynamic hypergame on graph model captures two important aspects about players' perceptions: (a) the evolving misperception of P2, and (b) the P1's information regarding current perception of P2. 

% On the contrary, we define a dynamic hypergame on graph model using a game on graph, instead of a normal form game. This allows us to adopt the existing algorithm for computing \ac{asw} region in a game on graph to propose an algorithm to compute \ac{dasw} region in a dynamic hypergame on graph. 

% The main contributions of this paper are the dynamic hypergame model and the algorithm to compute \ac{dasw} region. 

\paragraph*{A Modeling Framework with Dynamic Hypergame} The dynamic hypergame on graph  models (i) the evolving information of P2, and (ii) the P1's information regarding current perception of P2.

\paragraph*{\ac{dasw} Synthesis Algorithm} We propose an algorithm to identify the \ac{dasw} region and synthesize a \ac{dasw} strategy. We prove that the computed \ac{dasw} region is a superset of \ac{asw} region, which implies that the \ac{dasw} strategy is at least as powerful as the \ac{asw} strategy.

% =====================================================
% SECTION
% =====================================================
\section{Preliminaries}
Let $\Sigma$ be a finite alphabet. A sequence of symbols $w=w_0 w_1 \ldots w_n  $ with $w_i\in \Sigma, i=0,1,\ldots, n$ is called a \emph{finite word} and $\Sigma^\ast$ is the set of finite words that can be generated with alphabet $\Sigma$. We denote by $\Sigma^\omega$, the set of $\omega$-regular words obtained by concatenating the elements in $\Sigma$ infinitely many times. Given a set $X$, let $\dist{X}$ be the set of probability distributions over $X$. Given a distribution $\delta \in \dist{X}$, the set $\supp(\delta) = \{ x \in X \mid \delta(x) > 0\}$ is called the support of the distribution. 
% When $\supp(\delta)$ is singleton, the distribution $\delta$ is called as Dirac delta function. 

% =====================================================
% SUB-SECTION
% =====================================================
\subsection{Games on Graph} \label{subsect:game-on-graph}

Consider an interaction between two players; P1 with a reachability objective and P2 with an objective of preventing P1 from completing her task.

{\definition[Game on Graph] Let the action sets of P1 and P2 be $A_1$ and $A_2$, respectively. Then, a turn-based game on graph is the tuple \[\game = \langle S, \act, T, F \rangle ,\] where 

% Option 1: Paragraph Style Explanation of Terms
% $S = S_1 \cup S_2$ is the set of states partitioned into P1's states, $S_1$, and P2's states, $S_2$. P1 chooses an action when $s \in S_1$ and P2 chooses an action when $s \in S_2$. $\act = A \cup B$ is set of actions for P1 and P2. $T : (S_1 \times A) \cup (S_2 \times B) \rightarrow S$ is a deterministic transition function that given an P1's state and her action, or a P2's state and his action, gives the successor state. $F \subseteq S$ is a set of final states. 

% Option 2: Bullet Style Explanation of Terms
\begin{itemize}
    \item $S = S_1 \cup S_2$ is the set of states partitioned into P1's states, $S_1$, and P2's states, $S_2$. P1 chooses an action when $s \in S_1$ and P2 chooses an action when $s \in S_2$.
    
    \item $\act = A_1 \cup A_2$ is set of actions for P1 and P2.
    
    \item $T : S \times \act \rightarrow S$ is a deterministic transition function that maps a state and an action to a successor state.
    
    \item $F \subseteq S$ is a set of final states. 
\end{itemize} 
}

A \textit{trace} in the game $\game$ is an infinite, ordered sequence of state-action pairs $\tau = (s_0, a_0), (s_1, a_1), (s_2, a_2), \ldots $. We write $\tau[n] = (s_n, a_n)$ to denote $n$-th state-action pair, and $\tau[m:n] = (s_m, a_m), \ldots, (s_n, a_n)$ to denote a state-action pairs between $m$-th and $n$-th step, both inclusive. A \textit{run} is the projection of trace onto the state-space. We denote it as the sequence $\rho =  \tau \downharpoonright_S = s_0 s_1 s_2 \ldots$. Similarly, the \textit{action-history} is the projection of trace onto the action space, denoted by $\alpha = \tau \downharpoonright_\act   = a_0 a_1 a_2 \ldots$. The $k$-th element in a run (resp. action-history) is denoted by $\rho_k$ (resp. $\alpha_k$).

In this paper, we consider \textit{reachability objectives} for P1. The set of states that occur in a run is given by $\occ(\rho) = \{s \in S \mid \exists k \in \mathbb{N} \cdot s = \rho_k \}$. A run is said to be winning for P1 in the reachability objective if it satisfies $\occ(\rho) \cap F \ne \emptyset$. If a run is not winning for P1, then it is winning for P2.

\paragraph*{\ac{asw} Strategy} A stochastic or randomized strategies for P1 and P2 are defined as $\pi: S_1 \rightarrow \dist{A_1}$ and $\sigma: S_2 \rightarrow \dist{A_2}$, respectively. Let $\Omega_s^{\pi, \sigma}$ be the exhaustive set of runs that result when P1 and P2 play strategies $\pi$ and $\sigma$ in a game starting at the state $s \in S$. The randomized strategies of P1 and P2 induce a Markov chain from $\game$--that is, a probability distribution over the set $\Omega_s^{\pi, \sigma}$.

Given a state $s \in S$, a randomized strategy $\pi$ is \textit{almost-sure winning} for P1, if and only if for every possible randomized strategy $\sigma$ of P2, the probability is one for a run that satisfies $\occ(\rho) \cap F \neq \emptyset$, given the distribution of runs induced by $(\pi, \sigma)$. A state is called an \textit{almost-sure winning state} for P1, if there exists an almost-sure winning strategy for P1 from that state. The exhaustive set of almost-sure winning states for P1 is called her \textit{almost-sure winning region}. The almost-sure winning region can be computed using \alg{alg:zielonka} based on the \prop{lma:deAlfaro}.

% \textit{Sure-Winning Strategy:} Given a state $s \in S$, a deterministic strategy $\pi$ is said to be \textit{sure-winning} for P1 if and only if for every possible deterministic strategy $\sigma$ of P2, all runs $\rho$ corresponding to traces $\tau \in \Omega_s^{\pi, \sigma}$ are winning for P1. A state is a \textit{sure-winning state} for P1, if there exists a sure-winning strategy for P1 from that state. The exhaustive set of sure-winning states for P1 is called her \textit{sure-winning region}. The sure-winning region can be computed using the Zielonka's attractor algorithm \cite{zielonka1998} as given in \alg{alg:zielonka}.  

\begin{algorithm}
\caption{Zielonka's Algorithm}
\label{alg:zielonka}

\begin{algorithmic}[1]
\item[\textbf{Inputs:}] $\game, F$ 
    \State $Z_0 = F$
    \While{True}
        \State $\mathsf{Pre}_1(Z_k) = \{s \in S_1 \mid \exists a \in A_1 \text{ s.t. } T(s, a) \in Z_k \}$
        
        \State $\mathsf{Pre}_2(Z_k) = \{s \in S_2 \mid \forall b \in A_2: T(s, b) \in  Z_k\}$
        
        \State $Z_{k+1} = Z_k \cup \mathsf{Pre}_1(Z_k) \cup \mathsf{Pre}_2(Z_k)$
        \If {$Z_{k+1} = Z_k$}
            \State End Loop
        \EndIf
    \EndWhile
    \State \Return $Z_k$
\end{algorithmic}
\end{algorithm}

% \textit{Almost-Sure Winning Strategy: } Given a state $s \in S$, a stochastic strategy $\pi$ is \textit{almost-sure winning} for P1, if and only if for every possible stochastic strategy $\sigma$ of P2, the probability that every run $\rho$ corresponding to every trace $\tau \in \Omega_s^{\pi, \sigma}$ satisfies $\occ(\rho) \cap F \neq \emptyset$ is one. A state is called a \textit{almost-sure winning state} for P1, if there exists an almost-sure winning strategy for P1 from that state. The exhaustive set of almost-sure winning states for P1 is called her \textit{almost-sure winning region}. The almost-sure winning region may be computed using \alg{alg:zielonka} due to the following proposition. 

{\proposition(From \cite[Thm 3]{deAlfaro2007}) \label{lma:deAlfaro} In a deterministic and turn-based game, the almost-sure winning region is equal to sure-winning region. }

Let us introduce a running example that we shall use to explain the concepts in this paper.

{\example \label{ex:1} Consider the game graph in \fig{fig:game-1}. The circle states, $\{s_1, s_3\}$, are P1 states and the square states, $\{s_0, s_2\}$, are P2 states. The objective of P1 is to reach to the final state $s_0$ from the initial state $s_2$. P1's action set is $A_1 = \{a_1, a_2\}$, and P2's action set is $A_2 = \{b_1, b_2\}$.

\begin{figure}
    \centering
    \begin{tikzpicture}[->,>=stealth',shorten >=1pt,auto,node distance=3cm,scale=.75, semithick, transform shape,square/.style={regular polygon,regular polygon sides=4}]
    		\tikzstyle{every state}=[fill=black!10!white]
    		\node[accepting, state, square]     (0)                           {$s_0$};
    		\node[state]                        (1) [above of =0]             {$s_1$};
    		\node[initial above, state, square] (2) [right of =1]             {$s_2$};
    		\node[state]                        (3) [below of =2]             {$s_3$};
    		
    		\path[->]   
    		(1) edge                node{$a_1$}             (0)
            (1) edge[bend left]     node{$a_2$}             (2)
            (2) edge[bend left]     node[above]{$b_1$}      (1)
            (2) edge[bend right]    node[left]{$b_2$}       (3)
            (3) edge                node[right]{$a_1$}      (2)
            (3) edge[bend right]    node[right]{$a_2$}      (2)
           ;
    	\end{tikzpicture}
    \caption{An example game on graph}
    \label{fig:game-1}
\end{figure}
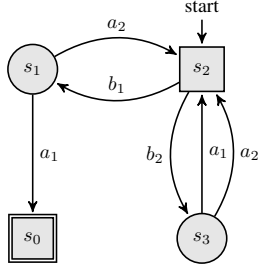

The \ac{asw} region for P1 in the game is $\win_1 = \{s_0, s_1\}$. This can intuitively be understood as follows. P1 can win from state $s_1$ by choosing the action $a_1$. However, the states $s_2$ and $s_3$ are losing for P1 because P2 has a strategy to indefinitely restrict the game within the states $s_2, s_3$ by choosing the action $b_2$ at the state $s_2$. 

}

% =====================================================
% SUB-SECTION
% =====================================================
\subsection{Action Misperception and Information Asymmetry}

% Recall that we are interested in two-player games under information asymmetry. In our problem setup, P1 has complete information about both players' action capabilities and P2 has incomplete information about P1's action capabilities, i.e. when P2 has misperception about P1's action set. We assume that both P1 and P2 have complete information about the states $S$, transition function $T$ and the final states $F$. 

In this paper, we make the following assumption about the reachability game, in which the P2 has asymmetric information about P1's action capabilities.

{\assumption \label{ass:subset} P1 has complete information about the players' action sets, \ie P1 knows $A_1$ and $A_2$. P2 only knows his own action set $A_2$, but (mis)perceives P1's action set to be a subset $X \subsetneq A_1$. Both players have complete information about the game state-space $S$, transition function $T$ and the final states $F$.
}

The result of \ass{ass:subset} is that P1 and P2, in their minds, play different games to synthesize their respective strategies. We refer to these games as the \textit{perceptual games} of the players. P1's  perceptual game is identical to the ground-truth game; $\game(A_1) = \langle S, A_1 \cup A_2, T, F \rangle$, while P2's perceptual game is a game under misperception; $\game(X) = \langle S, X \cup A_2, T, F \rangle$. Let us formalize the new notation used to distinguish between the perceptual games of P1 and P2.

{\notation \label{notation:game-with-action-set} Let $X \subseteq A_1$ be a subset of P1's action set. We denote a perceptual game in which P1's action set is $X$ by $\game(X) = \langle S, X \cup A_2, T, F \rangle$. The winning regions for P1 and P2 in the game $\game(X)$ are denoted by $\win_1(X)$ and $\win_2(X)$, respectively.  
}

Assuming P1 and P2 to be rational players, they would use the solution approach reviewed in \sect{subsect:game-on-graph} to compute their winning strategies in their respective perceptual games. That is, P1 will solve $\game(A_1)$ in her mind to obtain $\pi$ and P2 will solve $\game(X)$ in his mind to compute $\sigma$. However, P1 is likely to compute a conservative strategy; because she over-estimates the information available to P2. Naturally, we want to know 
\textit{whether P1 can improve her strategy if she is made aware of P2's current misperception $X$?}
% \textit{whether P1 can synthesize a less conservative winning strategy if she is aware of P2's current misperception, i.e. in addition to $A_1, A_2$ she also knows $X$?}

Before we answer the above question, recall from \sect{sec:introduction} that we allow P2's misperception to evolve during the game. For instance, what would happen when P2 observes P1 playing an action $a \in A_1$, which P2 did not believe to be in P1's action set? We might argue that P2 will at least add a new action $a$ to his perceived action set, $X$, of P1. Thus, the new perception would be $X \cup \{a\}$. Also, P2 might be capable of complex inference. That is, on observing that P1 can perform an action $a$, P2 might infer that P1 must be capable of actions $b$ and $c$, thus, updating his perception set to $X \cup \{a, b, c\}$. To capture such  inference capabilities, we introduce a generic perception update function for P2 as follows, 

% However, if P2 is more intelligent and is capable of inference, he might also infer more actions; say $b, c \in A_1$, that P1 can perform by simply observing $a$. Therefore, the newly updated perception with inference would be $X \cup \{a, b, c\}$. To capture such  inference capabilities, we introduce a generic action-set update function for P2 as follows, 

% In addition, P2 may also be capable of inference. For example, in a gridworld, if P2 observes P1 jumping 3 steps north, then he may infer that she is capable of jumping 2 steps to north. Therefore, on observing an action $a$, P2 might update $X$ to $X \cup \{a, b\}$, where $b \in A \setminus X$ is an inferred action. To capture such  inference capabilities, we introduce a generic action-set update function for P2 as follows, 

{\definition[Inference Mechanism] \label{def:inference-mechanism} A deterministic inference mechanism is a function $\update: 2^{A_1} \times A_1^\ast \rightarrow 2^{A_1}$ that maps a subset of actions $X \subseteq A_1$ and a finite action-history $\alpha$ to an updated subset of actions $Y = \update(X, \alpha)$ such that if there exists an action $a \notin X$ which is present in $\alpha$, then $a \in Y$. 
}

Given the formalism of inference mechanism to capture the evolving misperception of P2 during the game, we now proceed to defining our problem statement.

% =====================================================
% SUB-SECTION
% =====================================================
\subsection{Problem Statement}

When P2's misperception evolves during the game, P1 should also strategize to reveal an action that is not currently known to P2. By doing so, P1 may control the evolution of P2's misperception to her advantage. Let us revisit \ex{ex:1} to develop an intuition of how P1 might control P2's perception. 

{\example[\ex{ex:1} contd.] \label{ex:2} Suppose that, in \ex{ex:1}, P2 starts with a misperception about P1's action capabilities as $X_0 = \{a_2\}$. In this setup, let us understand the perceptual games of the players. P1's perceptual game; $\game_1 = \game(A_1)$, is the same as the ground-truth game as shown in \fig{fig:game-1}. P2's perceptual game, initially, is the game $\game_2 = \game(X_0)$ that does not include edges labeled with action $a_1$ as shown in \fig{fig:game-2}. Clearly, as the final state $s_0$ is not reachable in $\game_2$, P2 misperceives both actions $b_1$ and $b_2$ to be safe to play at state $s_2$, when only the action $b_2$ is safe in the ground-truth game.

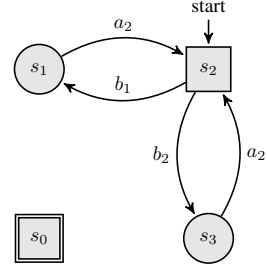
\begin{figure}
    \centering
    \begin{tikzpicture}[->,>=stealth',shorten >=1pt,auto,node distance=3cm,scale=.75, semithick, transform shape,square/.style={regular polygon,regular polygon sides=4}]
		\tikzstyle{every state}=[fill=black!10!white]
		\node[accepting, state, square]     (0)                           {$s_0$};
		\node[state]                        (1) [above of =0]             {$s_1$};
		\node[initial above, state, square] (2) [right of =1]             {$s_2$};
		\node[state]                        (3) [below of =2]             {$s_3$};
		
		\path[->]   
% 		(1) edge                node{$a_1$}             (0)
        (1) edge[bend left]     node{$a_2$}             (2)
        (2) edge[bend left]     node[above]{$b_1$}      (1)
        (2) edge[bend right]    node[left]{$b_2$}       (3)
        % (3) edge                node[right]{$a_1$}      (2)
        (3) edge[bend right]    node[right]{$a_2$}      (2)
       ;
	\end{tikzpicture}
    \caption{Perceptual game of P2}
    \label{fig:game-2}
\end{figure}

When P1 is aware of P2's misperception, $X_0$, a deceptive strategy should, intuitively, not use $a_1$ unless game state is $s_1$. Assuming P2 uses a randomized strategy with support $A_2$, it is easy to compute that the probability of reaching the state $s_1$ is one. At $s_1$, P1 can win the game by choosing $a_1$ in one step. We note that if P1 uses $a_1$ in state $s_3$, then P2 will update his perception to $X_1 = A_1$, and mark the action $b_1$ to be unsafe in state $s_2$. Thus, P1 will never be able to win the game. 
}

We call such a strategy of P1, where she intentionally controls P2's misperception, as an \textit{action-deceptive strategy} or simply a deceptive strategy (see \defn{defn:dec-almost-sure-win-strategy} for a formal definition). We formalize our problem statement.

{\problem \label{problem} Consider a reachability game under information asymmetry in which \ass{ass:subset} holds. If P1 is informed of the initial misperception of P2, $X_0$, and his inference mechanism $\eta$, then determine a \ac{dasw} strategy for P1 to satisfy her reachability objective.
}

In particular, we want to investigate whether the use of deception is advantageous for P1 or not. We say P1 gets advantage with deception if at least one game state that is almost-sure losing for P1 in the game without deception becomes winning for her with use of deception.

% =====================================================
% SECTION: DYNAMIC HYPERGAME OF ACTION DECEPTION
% =====================================================
\section{Dynamic Hypergame for Action Deception}

% A game on graph with reachability objectives can be represented as a two-player turn-based zero-sum game, with an assumption that both players have complete information and observation \cite{Buchi1969}. However, w
When two players play different games in their minds, their interaction is better modeled as a hypergame \cite{Bennett1977}.

{\definition[First-level Hypergame] A first-level hypergame is defined as a tuple of the perceptual games being played by the players, \[\hgame^1 = \langle \game_1, \game_2 \rangle, \]
where the P1 (resp. P2) solves the game $\game_1$ (resp. $\game_2$) to compute the winning strategy. 
}

When one of the players is aware of the other player's perception, but the other player is not, we say that a second-level hypergame is being played. In line with Problem~\ref{problem}, we assume that P1 is aware of the P2's misperception, \ie P1 knows the action set $X \subseteq A_1$ as perceived by P2. If P1 knows $X$, then P1 can construct the perceptual game of P2, $\game(X)$, and therefore P1 knows the first-level hypergame $\hgame^1$.

% If at least one of the players is aware of the other player's misperception; \ie one of the players knows the first-level hypergame $\hgame^1$ while the other player plays a perceptual game, then we say that a second-level hypergame is being played. In line with Problem~\ref{problem}, let us assume P1 knows $\hgame^1$ and P2 plays $\game(X)$. Therefore, in a second-level hypergame, P1 will compute her strategy by solving $\hgame^1$, while P2 will compute his strategy by solving $\game(X)$, where $X$ is P2's current perception of P1's action set. 

However, P2's perception evolves when he observes P1 using actions that are not included in $X$. This means that the game $\game(X)$ changes when P2's perception changes, and so does the hypergame $\hgame^1$. We now define a graph to model the hypergame representing the evolving misperception of P2, called as a dynamic hypergame on graph.

{\definition[Dynamic Hypergame on Graph] \label{defn:hypergame-graph} Let $\Gamma: \{1, 2, \ldots, 2^{|A_1|}\}$ be an indicator set. Let $\gamma: \Gamma \rightarrow 2^{A_1}$ be a bijection from the indicator set to the power set of $A_1$. We define the dynamic hypergame on graph as
\[
    \hgame = \langle V, \act, \Delta, \cal{F}  \rangle, 
\]
where 
\begin{itemize}
    \item $V = S \times \Gamma$  is the set of hypergame states, 
    
    \item $\act = A_1 \cup A_2$ is the set of actions of P1 and P2, 
    
    \item $\Delta: V \times \act \rightarrow V$ is the transition function such that $(s, i) \xrightarrow{a} (s', i')$ if and only if $s' = T(s, a)$ and $\gamma(i') = \update(X, \{a\})$ where $X = \gamma(i)$,
    
    \item ${\cal F} = F \times \Gamma$ is the set of final states. 
\end{itemize}
}

% {\remark The conventional games on graphs have memoryless winning strategy \cite{}. On the contrary, computing a action-deceptive winning strategy in dynamic hypergames will require P1 to maintain a finite memory of the actions she has already revealed to P2 and those which are still hidden from him. This requires a finite memory. To avoid dealing with finite-memory, we augment the state of P2's misperception to the state of the game to define a dynamic hypergame on graph.
% }

For convenience, we shall refer to the dynamic hypergame on graph as simply hypergame in the remainder of the paper. Analogous to game on graph, a trace in a hypergame is an infinite, ordered sequence of state-action pairs given by $\tau = (v_0, a_0) (v_1, a_1) \ldots$ and the action-history is defined as $\alpha = \tau \downharpoonright_\act  = a_0 a_1 a_2 \ldots$. In contrast with the game on graphs, we distinguish between a \textit{hypergame-run} (h-run) as a projection of trace onto the hypergame state-space $\nu = \tau \downharpoonright_V = v_0 v_1 v_2 \ldots $ and a \textit{game-run} as a projection of trace onto game state-space $\rho = \tau \downharpoonright_S = s_0 s_1 s_2 \ldots$, where $s_k$ is the game state corresponding to hypergame state $v_k = (s_k, \cdot)$. A reachability objective is said to be satisfied over the hypergame if and only if $\occ(\nu) \cap {\cal F} \neq \emptyset$, \ie the hypergame-run $\nu$ visits a winning state in $\cal F$. By definition, the following statement is always true; $\occ(\rho) \cap {F} \neq \emptyset$ if and only if $\occ(\nu) \cap {\cal F} \neq \emptyset$.

{\example[\ex{ex:1} contd.] \label{ex:idea-example-1} 
% Consider the game graph as shown in \fig{fig:idea-example-1a}. The circle states, $\{s_1, s_3\}$, are P1 states and square states,
% $\{s_0, s_2\}$, are P2 states. The objective of P1 is to reach to the final state $s_0$ from the initial state $s_2$. P1's action set is $A_1 = \{a_1, a_2\}$, and P2's action set is $A_2 = \{b_1, b_2\}$. P2's initial misperception of P1's action set is $X_0 = \{a_2\}$. 

% Given the above setup, let us look at perceptual games of the players. P1's perceptual game; $\game_1 = \game(A_1)$, is the same as the ground-truth game as shown in \fig{fig:idea-example-1a}. P2's perceptual game, initially, is the game $\game_2 = \game(X_0)$ that does not include edges labeled with action $a_1$ as shown in \fig{fig:idea-example-1b}. Clearly, as the final state $s_0$ is not reachable in $\game_2$, P2 misperceives both actions $b_1$ and $b_2$ to be safe to play at state $s_2$, when only the action $b_2$ is safe in ground-truth. 

% When P1 is aware of P2's misperception, $X_0$, a deceptive strategy should, intuitively, not use $a_1$ unless game state is $s_1$. Assuming P2 uses a randomized strategy with support $A_2$, we can say probability of reaching the state $s_1$ is one. At $s_1$, P1 can win the game by choosing $a_1$ in one step. We quickly note, if P1 uses $a_1$ in state $s_3$, then P2 will update his perception to $X_1 = A_1$, and mark the action $b_1$ to be unsafe in state $s_2$. Thus, P1 will never be able to win the game. 

The hypergame modeling the asymmetric information from \ex{ex:2} is shown in \fig{fig:hgame} (the figure only shows the reachable states). Every state is represented a tuple $(s_i, j)$ where $j = 0, 1$ represents the current misperception of P2. The bijection map is defined as $\gamma(0) = A_1$ and $\gamma(1) = \{a_2\}$. The traces $\tau_1 = \left((s_2, 1), b_1\right), \left((s_1, 1), a_1\right), (s_0, 0)$,  and $\tau_2 = \left((s_2, 1), b_2\right), \left((s_3, 1), a_2\right), \left((s_2, 0), b_1\right), \left((s_1, 1), a_1\right), (s_0, 0)$ are the examples of winning traces in the hypergame.

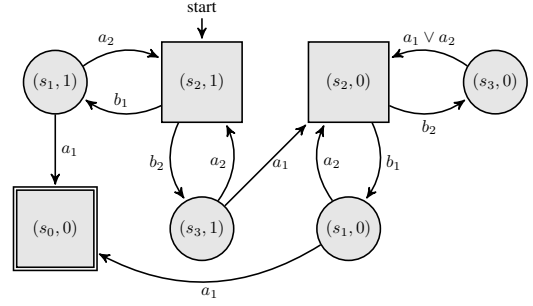
\begin{figure}
    \centering
    \begin{tikzpicture}[->,>=stealth',shorten >=1pt,auto,node distance=3cm,scale=.65,semithick, transform shape,square/.style={regular polygon,regular polygon sides=4}]
		\tikzstyle{every state}=[fill=black!10!white]
		\node[accepting, state, square]     (0)                           {$(s_0, 0)$};
		\node[state]                        (1) [above of =0]             {$(s_1, 1)$};
		\node[initial above, state, square] (2) [right of =1]             {$(s_2, 1)$};
		\node[state]                        (3) [below of =2]             {$(s_3, 1)$};
		\node[state]                        (4) [right of =3]             {$(s_1, 0)$};
		\node[state, square]                (5) [right of =2]             {$(s_2, 0)$};
		\node[state]                        (6) [right of =5]             {$(s_3, 0)$};
		
		\path[->]   
		(1) edge                node{$a_1$}                      (0)
        (1) edge[bend left]     node{$a_2$}                      (2)
        (2) edge[bend left]     node[above]{$b_1$}               (1)
        (2) edge[bend right]    node[left]{$b_2$}                (3)
        (3) edge                node[right]{$a_1$}               (5)
        (3) edge[bend right]    node[left]{$a_2$}                (2)
        (4) edge[bend left]     node[below]{$a_1$}               (0)
        (4) edge[bend left]     node[right]{$a_2$}               (5)
        (5) edge[bend left]     node[right]{$b_1$}               (4)
        (5) edge[bend right]    node[below]{$b_2$}               (6)
        (6) edge[bend right]    node[above]{$a_1 \lor a_2$}      (5)
       ;
	\end{tikzpicture}
    \caption{The dynamic hypergame on graph}
    \label{fig:hgame}
\end{figure}

% \todo[inline]{Move this para to next section.}
% Using the solution to game on graph from \sect{}, P1 will compute $\win_1(A) = \{s_0, s_1\}$, while P2 will compute $\win_1(X) = \{s_0\}$. Therefore, in state $s_2$, P2 may feel safe to play either $b_1$ or $b_2$. However, in state $s_3$ if P1 plays $a_1$, then P2 will update his misperception from $X$ to $A$ and, thereby, his strategy in state $s_2$ to playing only $b_2$. On the contrary, if P1 plays a deceptive strategy to ensure P2's misperception in $s_2$ is always $X$, then there is a possibility that P2 may play $b_1$, thus leading to a winning state in game graph. 

% The dynamic hypergame corresponding to above scenario is shown in \fig{fig:idea-example-1c}. The bijection mapping $\Gamma$ is defined as $\gamma(0) = \{a_1, a_2\}$ and $\gamma(1) = \{a_2\}$ in the figure. Therefore, we write a hypergame state $v = (s_3, 1)$ to represent the scenario when the game state is $s_3$ and P2's misperception is $\gamma(1) = \{a_2\}$. An example of a trace in $\hgame$ is $\tau = ((s_3, 1), a_2) ~((s_2, 1), b_2) ~((s_3, 1), a_2) ~((s_2, 1), b_1)$ $((s_1, 1), a_1) ~((s_0, 0), \emptyset)$. The run is winning because a final state $(s_0, 0)$ is in $\tau_V$. 
}

\section{Deceptive Almost-Sure Winning Strategy}

% Let us start by formalizing assumptions and defining the deceptive almost-sure winning strategy.

In this section, we present an algorithm to synthesize \ac{dasw} strategies in the hypergame. The algorithm relies on an assumption about P2's strategy, which requires us to revisit the concept of permissive strategies in a game on graph.

Recall that an action is permissive for a player at a given state if the player can stay within the winning region by performing that action \cite{bernet2002}. In a game under information asymmetry, whether a state is winning or not depends on the player's perception. Hence, we define the notion of perceptually permissive actions, which extends the definition of permissive actions to model evolving perception.

% In a game under information asymmetry and evolving misperception, the permissive strategy of P2 also keeps evolving with the perception. This notion is captured in the following definition. 

{\definition[Perceptually Permissive Actions] \label{defn:perceptually-permissive-strategy} Let $u = (s, i)$ and $v = (s', i')$ be two hypergame states such that $v = \Delta(u, b)$ for some $b \in A_2$. Let $X = \gamma(i)$ and $X' = \gamma(i')$ be the misperception of P2 at states $u$ and $v$, respectively. Then, the set $M(u) = \{b \in A_2 \mid s' \in \win_2(X')\}$ is the set of perceptually permissive actions at $u$. 

% A \ac{rpps} is a distribution $\mu(u) = \dist{M(u)}$, such that the support of the $\mu$ is $M(u)$. 

% Given a state $v = (s, i)$ in hypergame $\hgame$, the set of perceptually permissive actions for P2 at $v$ is $M(v) = \{b \in A_2 \mid (s', i') = \Delta(v, b) \text{ and } s' \in \win_2(\gamma(i')) \}$. A perceptually permissive strategy of Bob is defined over the set of perceptually permissive actions of Bob as a map $\mu: V_2 \rightarrow \dist{A_2}$ such that $\supp(\mu(v)) = M(v)$. 
}
In words, the perceptual permissive actions for a given state $u=(s,i)$ is the set of permissive actions in the perceptual game, with index $i$.

{\assumption \label{ass:stochastic-strategy} At a state $v \in V_2$, P2 plays a randomized strategy, $\sigma$, defined over the perceptually permissive actions $M(v)$ such that $\supp\left(\sigma(v)\right) = M(v)$. 
} \vspace{0.5em}

Now, we formalize the notion of \ac{dasw} strategy. 

% The \ass{ass:stochastic-strategy} has two important implications. First, at a state $(s, i) \in V$, P2 may choose any one of his action that he believes to be safe for himself at the state $s \in S$, given his current misperception $X = \gamma(i)$. Secondly, as the probability selecting a perceptually permissive action is strictly positive, the probability of reaching to different perceptually safe states from $(s, i)$ is also positive. 

% The idea behind the \ass{ass:stochastic-strategy} is to capture the notion that, at a state $(s, i) \in V$, P2 may choose any one of his action that he believes to be safe for himself at the state $s \in S$, given his current misperception $X = \gamma(i)$. The randomness in his strategy generates a possibility that P2 may choose a different action at the same state when it is revisited, just as we saw in \ex{ex:idea-example-1}. 
% This mitigates one of the main causes of conservativeness of the deceptive sure-winning strategy. 

{\definition[Deceptive Almost-Sure Winning Strategy] \label{defn:dec-almost-sure-win-strategy} Given a hypergame state $v \in V$, a strategy $\pi$ is said to be \textit{deceptive almost-sure winning strategy} for P1 if and only if for every strategy of P2 satisfying \ass{ass:stochastic-strategy}, the probability of  an h-run $\nu$ induced from $\hgame$ by $(\pi, \sigma)$ satisfying $\occ(\nu) \cap {\cal F} \neq \emptyset$ is one. 
}

The states at which P1 has a \ac{dasw} strategy are called as \ac{dasw} states. The exhaustive set of all \ac{dasw} states is called as \ac{dasw} region.

Now, we discuss \alg{alg:DASW} that computes the \ac{dasw} region for P1. Our algorithm is inspired by the algorithm presented in  \cite{deAlfaro2007} to compute the \ac{asw} region in the concurrent $\omega$-regular games. The idea behind \alg{alg:DASW} is to identify the states where P2 perceives some unsafe actions as safe due to misperception. This is achieved by modifying the \textsc{Safe-1} and \textsc{Safe-2} sub-routines from \ac{asw} region computation algorithm in \cite{deAlfaro2007} using the following definitions: 
\begin{align*}
    \dapre_1^1(U) &= \{v \in V_1 \mid \exists a \in A \text{ s.t. } \Delta(v, a) \in U\}, \\ 
    \dapre_1^2(U) &= \{v \in V_2 \mid \forall b \in M(v) \text{ s.t. } \Delta(v, b) \in U\} ,\\
    \dapre_2^1(U) &= \{v \in V_1 \mid \forall a \in A \text{ s.t. } \Delta(v, a) \in U\}, \\
    \dapre_2^2(U) &= \{v \in V_2 \mid \forall b \in M(v) \text{ s.t. } \Delta(v, b) \in U\}.
\end{align*}

The \alg{alg:DASW} works as follows. It starts with the \ac{asw} region; $Z_0 = \win_1(A_1) \times \Gamma$, and then iteratively expands it  by invoking \textsc{Safe-2} followed by \textsc{Safe-1} until a fixed-point is reached. The \textsc{Safe-1} sub-routine computes the largest subset $U$ of the input set $Y$, such that P1 has a strategy to restrict the game indefinitely within $U$. \textsc{Safe-2} sub-routine computes the largest subset $U$ of the input set $Y$, such that P2; given his current (mis)perception, can restrict the game indefinitely within $U$. Here, it is important to note that P2 chooses his actions based on his perceptual game $\game(X)$, and not the hypergame. Only P1 knows the hypergame because she is aware of P2's misperception. As a consequence, before reaching the fixed-point,  \textsc{Safe-2} might include states from which P2 may not have a strategy to indefinitely restrict P1 from reaching $Z_0$, \ie P1 may have a \ac{dasw} strategy from these states. However, after reaching the fixed-point, say in the iteration $k$, we show that all \ac{dasw} states are included in $Z_k$. A \ac{dasw} strategy can then be computed based on the proof of \thm{thm:DASW-sufficiency}. Let us now revisit the \ex{ex:idea-example-1} to understand \alg{alg:DASW}.

\begin{algorithm}
\caption{Computation of the \ac{dasw} region and strategy for P1}
\label{alg:DASW}
\begin{algorithmic}[1]
    \Function{DASW}{$\hgame$}
        \State $Z_0 = \win_1(A) \times \Gamma$
        
        \While{True}
            \State $C_k = \textsc{Safe-2}(V \setminus Z_k)$ 
            
            \State $Z_{k+1} = \textsc{Safe-1}(V \setminus C_k)$ 
            
            \If{$Z_{k+1} = Z_{k}$}
                \State End loop
            \EndIf
        \EndWhile
        
        \State \Return $Z_k$
    \EndFunction
\end{algorithmic}
\vspace{0.5em}
\begin{algorithmic}[1]
    \Function{\textsc{Safe-$i$}}{$U$}
        \State $Y_0 = U$
        
        \While{True} 
            \State $W_1 = \dapre_i^1(Y_k)$
            \State $W_2 = \dapre_i^2(Y_k)$
            
            \State $Y_{k+1} = Y_k \cap (W_1 \cup  W_2)$
            
            \If{$Y_{k+1} = Y_{k}$}
                \State End loop
            \EndIf
        \EndWhile
        \State \Return{$Y_k$}
    \EndFunction
\end{algorithmic}

% \begin{algorithmic}[1]
%     \Procedure{Safe-2}{$Z$}
%     \State $U_0 = Z$
        
%         \While{True} 
%             \State $\mathsf{DAPre_2^1} = \{v \in V_1 \mid \forall a \in A \text{ s.t. } \Delta(v, a) \in Z\}$
            
%             \State $\mathsf{DAPre_2^2} = \{v \in V_2 \mid \forall b \in \mu(v) \text{ s.t. } \Delta(v, b) \in Z\}$
            
%             \State $U_{k+1} = U_k \cap (\mathsf{DAPre_2^1} \cup  \mathsf{DAPre_2^1})$
            
%             \If{$U_{k+1} = U_{k}$}
%                 \State End loop
%             \EndIf
%         \EndWhile
%     \EndProcedure
% \end{algorithmic}
% \vspace{0.5em}

\end{algorithm}

{\example[\ex{ex:idea-example-1} contd.] \label{ex:4} Consider the hypergame graph as shown in \fig{fig:hgame}. Recall from \ex{ex:1} that \ac{asw} region is $\win_1(A_1) = \{s_0, s_1\}$, therefore, we have $Z_0 = \{(s_0, 0), (s_1, 0), (s_1, 1)\}$. The perceptually permissive actions for P2 are $\mu((s_2, 1)) = \{b_1, b_2\}$ and $\mu((s_2, 0)) = \{b_2\}$.

\textbf{Iteration 1 of \textsc{DASW}.} The first step is to compute $C_0$, \ie the subset of $V \setminus Z_0$ which P2 perceives to be safe for himself. The \textsc{Safe-2} sub-routine takes 3 iterations to reach a fixed-point, at the end of which $C_0 = \{(s_2, 0), (s_3, 0)\}$. The next step is to compute $Z_1$, which the largest subset of $V \setminus C_0$ in which P1 can stay indefinitely. The \textsc{Safe-1} sub-routine takes 2 iterations to reach a fixed point. In its first iteration, $\dapre_1^1$ adds a state $(s_3, 1)$ and $\dapre_1^2$ adds a state $(s_2, 1)$ to $Z_1$. The interesting observation here is that $(s_2, 1)$ is added because the actions $b_1$ and $b_2$ are perceptually permissive actions for P2, both of which lead to a state in $V \setminus C_0$. 

\textbf{Iteration 2 of \textsc{DASW}.} The fixed-point of \textsc{DASW} algorithm is reached in second iteration with $Z_2 = \{(s_0, 0), (s_1, 0), (s_1, 1), (s_2, 1), (s_3, 1)\}$. The states $(s_2, 1)$ and $(s_3, 1)$ are idenitifed as the \ac{dasw} states for P1. 
}

With this intuition, we proceed to proving our first main result that establishes the existence of a game state $s \in S$ which is losing for P1 in the game $\game(A_1)$, but becomes winning for P1 in the hypergame by using action-deception.

{\theorem \label{thm:DASW-not-empty} The \ac{dasw} region may contain a state $v = (s, i)$ such that $s \notin \win_1(A_1)$. 

\proof We want to show the existence of an example where a hypergame state $(s, i) \in V$ is a \ac{dasw} state but the game state $s$ is not \ac{asw} state for P1. Observe that the states $(s_2, 1)$ and $(s_3, 1)$ in \ex{ex:4} satisfy the above condition. \qed 

}

Next, we proceed to prove the correctness of \alg{alg:DASW} by showing that from every state identified by the algorithm as a \ac{dasw} state, we can construct a \ac{dasw} strategy for P1 to ensure a visit to final states with probability one. We first prove two lemmas.

{\lemma \label{lma:stay-in-strategy} In the $i$-th iteration of \alg{alg:DASW}, P1 has a strategy to restrict the game indefinitely within $Z_i$, for all states in $Z_i$. 

% At every state $v \in V$ added to $Z_i$ in iteration $i \geq 0$ of \alg{alg:DASW}, P1 has a strategy to restrict the game within $Z_i$ indefinitely.

\proof ($v \in V_2$). For a P2's state in $Z_i$, every state $v' = \Delta(v, b)$ for a perceptually permissive action $b \in \mu(v)$ of P2 is in $Z_i$, by definition of $\mathsf{DAPre}_1^2$. Hence, no action of P2 at any state $v \in Z_i$ can lead the game state outside $Z_i$. 

($v \in V_1$). For every P1's state in $Z_i$, there exists an action $a \in A$ such that the successor $v' = \Delta(v, a)$ is in $Z_i$, by definition of $\mathsf{DAPre}_1^1$. Hence, P1 always has an action, consequently a strategy, to stay within $Z_i$. \qed 
}

{\lemma \label{lma:progressive-action} For every state $v \in Z_{i+1} \setminus Z_{i}$ added in the $i$-th iteration of \alg{alg:DASW}, there exists an action that leads into $Z_{i}$.
% from every state $v \in Z_{i+1} \setminus Z_{i}$.

\proof Consider the partitions of $V$ at the beginning of the $i$-th iteration. There can be at most 3 partitions; namely (a) $C_{i-1}$, (b) $Z_i$, and (c) $V \setminus (C_{i-1} \cup Z_i)$. We will prove the statement by showing that the every new state added to $Z_{i+1}$ has at least one transition into $Z_i$.

Consider $i$-th iteration of \alg{alg:DASW}. The sub-routine \textsc{Safe-2} will add a P1 state $v \in V_1 \setminus Z_i$ to $C_i$ if all the actions of P1 stay within $V \setminus Z_i$. Similarly, \textsc{Safe-2} will include a P2 state $v \in V_2 \setminus Z_i$ in $C_i$ if all perceptually permissive actions of P2 lead to a state within $V \setminus Z_i$. Therefore, a state that is not included in $C_i$ must have at least one action leading outside $V \setminus Z_i$, \ie entering $Z_i$. In the next step, the sub-routine \textsc{Safe-1} may add new states to $Z_{i+1}$ from the set $V \setminus C_i$. But, all states in $V \setminus C_i$ have an action entering $Z_i$. Hence, all new states added to $Z_{i+1}$ satisfy the statement. \qed

}

From \lma{lma:progressive-action}, it is easy to see that P1 has a strategy to reach $Z_i$ from a state added to $Z_{i+1}$ in one-step. However, this is not true for P2. From a P2 state in $Z_{i+1}$, there exists a positive probability to reach $Z_i$ because of \ass{ass:stochastic-strategy}. In the next theorem, we prove a stronger statement that from every state in $Z_{i+1}$, P1 can reach not only $Z_i$ but also $Z_0$ with probability one.

{\theorem \label{thm:DASW-sufficiency} From every \ac{dasw} state, P1 has a \ac{dasw} strategy. 

\proof The proof follows from \lma{lma:stay-in-strategy} and \lma{lma:progressive-action}. For any $v \in Z_i$, P1 has a strategy to stay within $Z_i$ indefinitely, by \lma{lma:stay-in-strategy}. Furthermore, by \lma{lma:progressive-action}, the probability of reaching to a state $v' \in Z_{i-1}$ from $v$ is strictly positive. Therefore, given a run of infinite length, the probability of reaching $Z_{i-1}$ from $Z_i$ is one. By repeatedly applying the above argument, we have the probability of reaching $Z_0$ from $Z_i$ is one. \qed
}

The \ac{dasw} strategy can be constructed based on the proof of \thm{thm:DASW-sufficiency}. At a P1 state $v \in V_1$, if $i \geq 1$ is the smallest integer such that $v \in Z_i$, then $\pi(v) = \{a \in A_1 \mid v' = \Delta(v, a) \text{ and } v' \in Z_{i-1}\}$ is the \ac{dasw} strategy of P1 at $v$. We also state the following two important corollaries (proofs omitted due to space) follow from \thm{thm:DASW-not-empty} and  \lma{lma:progressive-action}. 

{\corollary For every $i \geq 0$, we have $Z_i \subseteq Z_{i+1}$.
}

{\corollary The projection of \ac{dasw} region onto the game states is a superset of the \ac{asw} region.  
}

% {\lemma \label{lma:Z-containment} In \alg{alg:DASW}, $Z_k \subseteq Z_{k+1}$ for all $k \geq 0$. 
% }

% {\corollary \label{cor:strategy-to-stay} At a state $v \in V_1 \cap Z_k$, for some $k \geq 0$, P1 has a strategy to stay within $Z_k$ indefinitely.
% }

% =====================================================
% SECTION
% =====================================================
\section{An Illustrative Example}

\begin{figure}
    \centering
    \includegraphics[scale=0.5]{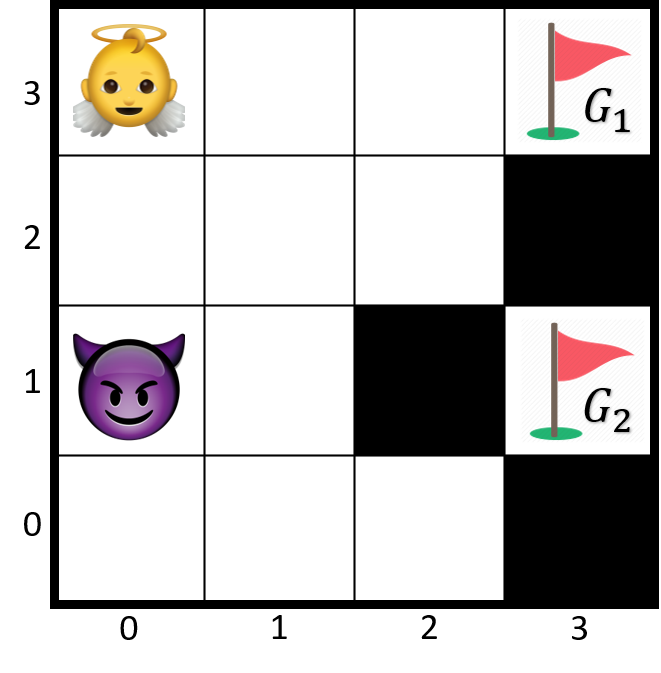}
    \caption{A game between a robot and its adversary on the gridworld.}
    \label{fig:illustration}
\end{figure}

We present a robot motion planning example over a $4 \times 4$ gridworld, shown in \fig{fig:illustration}, to illustrate how a robot (P1) may use action deception in presence of an adversary (P2). The objective of the robot is to visit the two cells $(3, 1)$ and $(3, 3)$ containing the flags, while the task of the adversary is to prevent this. The readers familiar with \ac{ltl} may recognize the above objective as a co-safe \ac{ltl} specification $\Eventually G_1 \land \Eventually G_2$.  The action set of the robot is $A_1 = \{\text{N, E, S, W, NE, NW, SW}\}$ while that of adversary is $A_2 = \{\text{N, E, S, W}\}$, where N, E, S, W stand for north, east, south and west. At the start of the game, the adversary has incomplete information about the robot's action set as $X_0 = \{\text{N, E, S, W}\}$. When the adversary observes the robot performing any of the actions from $\{$NE, NW, SW$\}$, he updates his perception to $X_1 = A_1$.

% \todo[inline]{What is the task of P1?}

A game on graph representing above scenario can be constructed using the product operation given in \cite[Def. 4.16]{Baier2008}. Every game state is a tuple $(x_1, y_1, x_2, y_2, t, q)$ where $x_i, y_i$ for $i=1,2$ denote the cell that P1 and P2 occupy, $t$ represents the player who chooses the next move and $q$ denotes a state of a \ac{dfa} that keeps track of the progress P1 has made towards completion of her objective. The resulting game has $4^4 \times 2 \times 4 = 2048$ states. We mark the states where P1 or P2 collide with an obstacle or with each other as the losing states for both players and, therefore, any action that leads to such states is disabled. Given the game on graph, a hypergame graph is constructed according to \defn{defn:hypergame-graph}. The hypergame graph has $2048 \times 2 = 4096$ states because the adversary has two information states; $X_0$ and $X_1$.

When the above hypergame graph is given as an input to the \alg{alg:DASW}, $2106$  out of $4096$ states are identified as \ac{dasw} states. The projection of \ac{dasw} states onto game state space results in $1172$ states, while the \ac{asw} region has the size of $934$ states. This means that $1172 - 934 = 238$ game states that were almost-sure losing for P1 became winning for her, when P1 incorporated P2's misperception into her planning to synthesize a deceptive strategy.

% =====================================================
% SECTION
% =====================================================
\section{Conclusion}

In this paper, we introduce a hypergame model to represent the interactions between two players with asymmetric information about their action capabilities. Then, we present an algorithm to synthesize action-deceptive strategies in a two-player turn-based zero-sum reachability games, where P2 has incomplete information about the P1's action capabilities. We show that the synthesized strategy has two important and desirable properties. First, the \ac{dasw} strategy is guaranteed to satisfy the reachability objective with probability one. Second, it is at least as powerful as the \ac{asw} strategy, because the \ac{dasw} region is a superset of the \ac{asw} region. 
% And third, the synthesis algorithm runs in polynomial-time in the size of the hypergame graph. 

% =====================================================
% BIBLIOGRAPHY
% =====================================================
%  Includes only that what is cited! 
\bibliographystyle{named}
\bibliography{action-deception}

\end{document}